\documentclass[conference]{IEEEtran}
\IEEEoverridecommandlockouts
\usepackage{cite}
\usepackage{amsmath,amssymb,amsfonts}
\usepackage{algorithmic}
\usepackage{graphicx}
\usepackage{textcomp}
\usepackage{xcolor}
\usepackage{textgreek}
\usepackage{graphicx}
\usepackage{subfigure}
\usepackage{float}
\usepackage{booktabs}  
\usepackage{amsmath}    
\usepackage{amssymb}    

\def\BibTeX{{\rm B\kern-.05em{\sc i\kern-.025em b}\kern-.08em
    T\kern-.1667em\lower.7ex\hbox{E}\kern-.125emX}}

\usepackage[bottom=1.02in,top=0.71in,left=0.7in,right=0.7in]{geometry}
\begin{document}

\title{SALT-V: Lightweight Authentication for 5G V2X Broadcasting
\vspace{-0.1cm}
}


\author{
    \IEEEauthorblockN{ 
        Liu Cao\IEEEauthorrefmark{1}, Weizheng Wang\IEEEauthorrefmark{2}, Qipeng Xie\IEEEauthorrefmark{3},
        Dongyu Wei\IEEEauthorrefmark{4},
        Lyutianyang Zhang\IEEEauthorrefmark{5}\\
    }
    \IEEEauthorblockA{
        \IEEEauthorrefmark{1}Department of Electronic Information Engineering, City University of Hong Kong (Dongguan), Dongguan, China\\
        \IEEEauthorrefmark{2}Department of Computer Science, City University of Hong Kong, Hong Kong SAR\\
        \IEEEauthorrefmark{3}Information Hub, Hong Kong University of Science and Technology (Guangzhou), Guangzhou, China\\
         \IEEEauthorrefmark{4} Department of Electrical and Computer Engineering, University of Miami, Coral Gables, FL, 33146, USA\\
         \IEEEauthorrefmark{5}The School of Microelectronics and Communication Engineering, Chongqing University, Chongqing, China\\
        Emails: liu.cao@cityu-dg.edu.cn, weizheng.wang@ieee.org, qxieaf@connect.ust.hk, \\dongyu.wei@miami.edu,
        zhanglyutianyang@cqu.edu.cn
        \vspace{-0.5cm}
    }

    \thanks{This work was supported by the Youth Innovation Talent Project of Guangdong Provincial Universities (Grant No. 2025KQNCX17).}
}

\maketitle

\begin{abstract}
Vehicle-to-Everything (V2X) communication faces a critical authentication dilemma: traditional public-key schemes like ECDSA provide strong security but impose 2 ms verification delays unsuitable for collision avoidance, while symmetric approaches like TESLA achieve microsecond-level efficiency at the cost of 20-100 ms key disclosure latency. Neither meets 5G New Radio (NR)-V2X's stringent requirements for both immediate authentication and computational efficiency. This paper presents SALT-V, a novel hybrid authentication framework that reconciles this fundamental trade-off through intelligent protocol stratification. SALT-V employs ECDSA signatures for 10\% of traffic (BOOT frames) to establish sender trust, then leverages this trust anchor to authenticate 90\% of messages (DATA frames) using lightweight GMAC operations. The core innovation—an Ephemeral Session Tag (EST) whitelist mechanism—enables 95\% of messages to achieve immediate verification without waiting for key disclosure, while Bloom filter integration provides O(1) revocation checking in 1 μs. Comprehensive evaluation demonstrates that SALT-V achieves 0.035 ms average computation time (57× faster than pure ECDSA), 1 ms end-to-end latency, 41-byte overhead, and linear scalability to 2000 vehicles, making it the first practical solution to satisfy all safety-critical requirements for real-time V2X deployment.

\end{abstract}


\section{Introduction}
\label{sec:introduction}

The deployment of Vehicle-to-Everything (V2X) communication promises to revolutionize transportation safety by enabling vehicles to exchange real-time information about their position, velocity, and intentions \cite{huang2025vehicle}. Through direct vehicle-to-vehicle (V2V) and vehicle-to-infrastructure (V2I) communication, V2X systems can prevent collisions, optimize traffic flow, and support autonomous driving. The U.S. Department of Transportation estimates that V2X deployment could eliminate or mitigate up to 80\% of non-impaired crashes \cite{Huang2022JSR}. However, this life-saving potential depends critically on the authenticity of exchanged messages—a single forged emergency brake warning could trigger multi-vehicle pileups, while fake congestion reports could gridlock entire cities \cite{chu2024survey}.

The 5G New Radio V2X (NR-V2X) standard demands unprecedented performance for authentication: messages must be verified within 100 ms to enable collision avoidance, while vehicles broadcast at 10 Hz, generating tens of thousands per second in dense urban areas \cite{Garcia2021NRV2X}. This creates a fundamental tension between security and efficiency. Traditional public key infrastructure (PKI) solutions, such as IEEE 1609.2, require 2 ms to verify each ECDSA signature—consuming significant computational resources and draining battery in electric vehicles \cite{Javed2016Sensors,Haas2009Globecom}. Conversely, symmetric-key alternatives like TESLA achieve microsecond-level performance but require receivers to wait for delayed key disclosure, typically 20-100 ms, before authenticating messages \cite{perrig2003tesla,Studer2009JCN,cao2022resource,cao2022optimize}. For emergency brake warnings where every millisecond counts, such delays are unacceptable.

Existing hybrid approaches attempt to balance these trade-offs but fall short of practical requirements. VAST \cite{studer2009tacking,Lyu2016TDSC} combines periodic signatures with TESLA but achieves only 1\% immediate verification rate. ECDSA-based schemes \cite{Cominetti2023VehCom} that reduce signature frequency sacrifice security during non-signed periods . Other proposals require additional infrastructure, trusted hardware, or complex group management unsuitable for dynamic vehicular networks. The challenge remains: how can we achieve both sub-millisecond computational efficiency and immediate verification capability without compromising security?

This paper presents SALT-V (Slot-Attested Lightweight TESLA for V2X), a novel authentication framework that resolves this dilemma through intelligent protocol design. Our key insight is that trust establishment and message authentication can be temporally decoupled: vehicles periodically broadcast BOOT frames (10\% of traffic) containing digital signatures to establish sender authenticity, then leverage this trust to authenticate subsequent DATA frames (90\% of traffic) using lightweight symmetric cryptography. The critical innovation lies in our Ephemeral Session Tag (EST) mechanism, which maintains a temporary whitelist of verified senders, enabling immediate authentication for 95\% of messages without waiting for key disclosure.

The primary contributions of this paper are listed as follows:
\begin{enumerate}
\item \textbf{Hybrid Adaptive Architecture}: A dual-frame structure where BOOT frames (10\% of traffic) use ECDSA signatures for trust establishment, while DATA frames (90\%) employ GMAC for efficiency, achieving 0.035 ms average computation—57× faster than pure signatures \cite{Cominetti2023VehCom}.

\item \textbf{Whitelist-based Immediate Verification}: The EST mechanism enables 95\% immediate authentication by maintaining short-lived trust relationships, eliminating TESLA's mandatory key disclosure delay for trusted senders.

\item \textbf{Efficient Revocation System}: Bloom filter integration provides O(1) revocation checking in 1 μs—10,000× faster than CRLs—handling millions of revocations in under 4 MB with 0.1\% false positive rate \cite{Jin2020BF,Broder2004IM}.

\item \textbf{Comprehensive Evaluation}: Extensive performance analysis across 2000 vehicle scenarios demonstrates practical deployment readiness with 341-byte messages, 1ms average delay, and proven scalability, validated through formal security analysis against replay, forgery, and privacy attacks.
\end{enumerate}

\section{Related Work}
\label{sec:related}

V2X authentication research has evolved along three primary trajectories: per-message signatures offering strong security but prohibitive overhead, TESLA-based schemes providing efficiency but lacking immediate verification, and hybrid solutions attempting to balance both but achieving limited success \cite{perrig2003tesla}. Traditional V2X security frameworks, exemplified by IEEE 1609.2 and DSRC/WAVE standards \cite{cao2020performance}, authenticate every message using ECDSA signatures with short-lived pseudonyms managed by Vehicular PKI systems \cite{sharma2020security}. While these approaches ensure strong source authentication and non-repudiation, verification overhead scales linearly with message rate, creating computational bottlenecks incompatible with 5G NR-V2X's millisecond-scale requirements \cite{Kenney2011DSRC,Brecht2018SCMS}. Various optimizations including batch ECDSA verification, ECQV implicit certificates, and reduced signature frequency schemes have been proposed to reduce average costs, yet these improvements only address constant factors and fail under worst-case scenarios such as broadcast storms or DoS attacks \cite{Zhang2008Batch,Pollicino2020ECQV}.

TESLA and its variants represent a fundamentally different approach, achieving minimal computational overhead through delayed key disclosure: senders commit to MAC keys and reveal them after a predetermined delay, enabling retrospective message verification using only symmetric primitives. Despite excellent scalability and efficiency, the mandatory disclosure delay—typically 20-50 ms—fundamentally conflicts with emergency applications requiring immediate authentication \cite{RFC4082TESLA}. Hybrid schemes like VAST attempt to bridge this gap by combining periodic signatures as trust anchors with TESLA for intervening messages, yet immediate verification remains confined to the sparse signature frames, typically under 5\% of traffic \cite{Studer2009VAST}. Complementary approaches leverage prediction or physical-layer information, such as PBA's kinematic prediction to reduce signature frequency, though robustness under adverse conditions remains challenging \cite{Lyu2016PBA}. Recent advances in revocation management using Bloom filters enable constant-time CRL checks but require careful integration with authentication mechanisms to prevent attacks during transition windows \cite{Khodaei2018CRL}.

Against this landscape, SALT-V advances the state-of-the-art through a refined hybrid architecture that achieves both efficiency and immediacy: signatures are minimized to 10\% of traffic (BOOT frames) for trust establishment, while 90\% of messages use lightweight GMAC authentication. The key innovation lies in the Ephemeral Session Tag mechanism that maintains a whitelist of verified senders, enabling 95\% immediate verification without disclosure delays. Combined with integrated Bloom filter revocation, SALT-V achieves 0.035 ms average computation—57× faster than pure signatures—with only 1 ms average latency, providing the first practical solution meeting all requirements for real-time V2X deployment.

\begin{figure}[t]
\centering
\includegraphics[width=0.8\columnwidth]{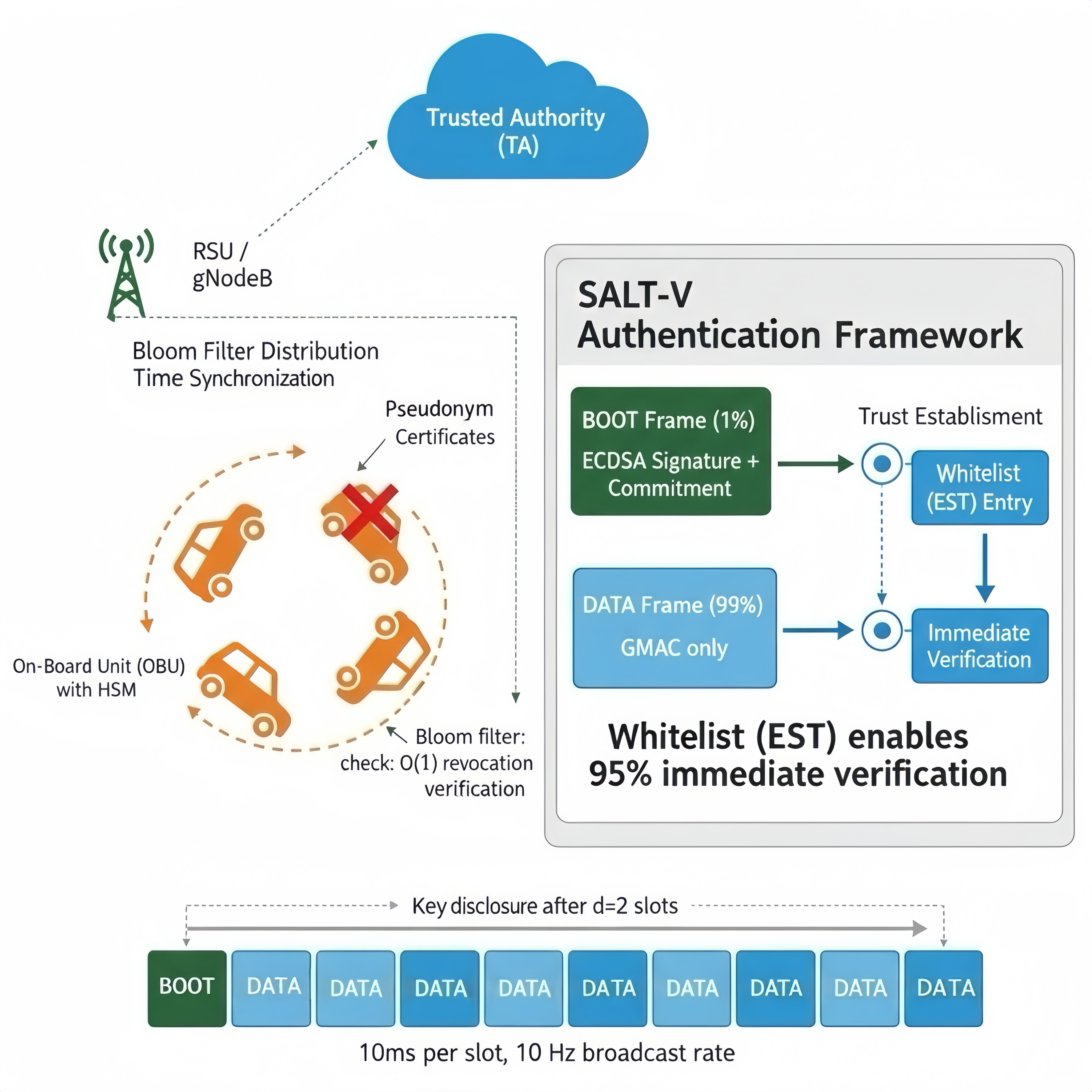}
\vspace{-0.4cm}
\caption{SALT-V system architecture and authentication flow}
\label{fig:system}
\vspace{-0.7cm}
\end{figure}

\section{System model}
\label{sec:System}
This section describes the network architecture, communication model, and security assumptions underlying the SALT-V authentication framework, as illustrated in Fig~\ref{fig:system}.

\subsection{Network Architecture}
The 5G NR-V2X system comprises three entities: (1) Trusted Authority (TA) for initialization and credentials, (2) Roadside Units (RSUs/gNodeBs) for infrastructure and time synchronization, (3) On-Board Units (OBUs) broadcasting safety messages. Vehicles operate in direct mode (PC5) broadcasting BSMs at 10 Hz within 300m, containing position, velocity, acceleration, and heading data for collision avoidance.

\subsection{Communication Model}
V2X employs one-to-many broadcasting without acknowledgment. Time is slotted ($T_s = 10$ms) and synchronized via RSUs. Vehicles derive authentication keys from a master seed per epoch (hourly). Messages include BOOT frames (10\%, 1 Hz) with signatures for trust establishment and DATA frames (90\%) with symmetric authentication. The system tolerates packet loss without ordering requirements \cite{11096953}.

\section{Our Proposed Scheme: SALT-V}
\label{sec:proposed}
\begin{figure}[t]
    \centering
    \includegraphics[scale = 0.30]{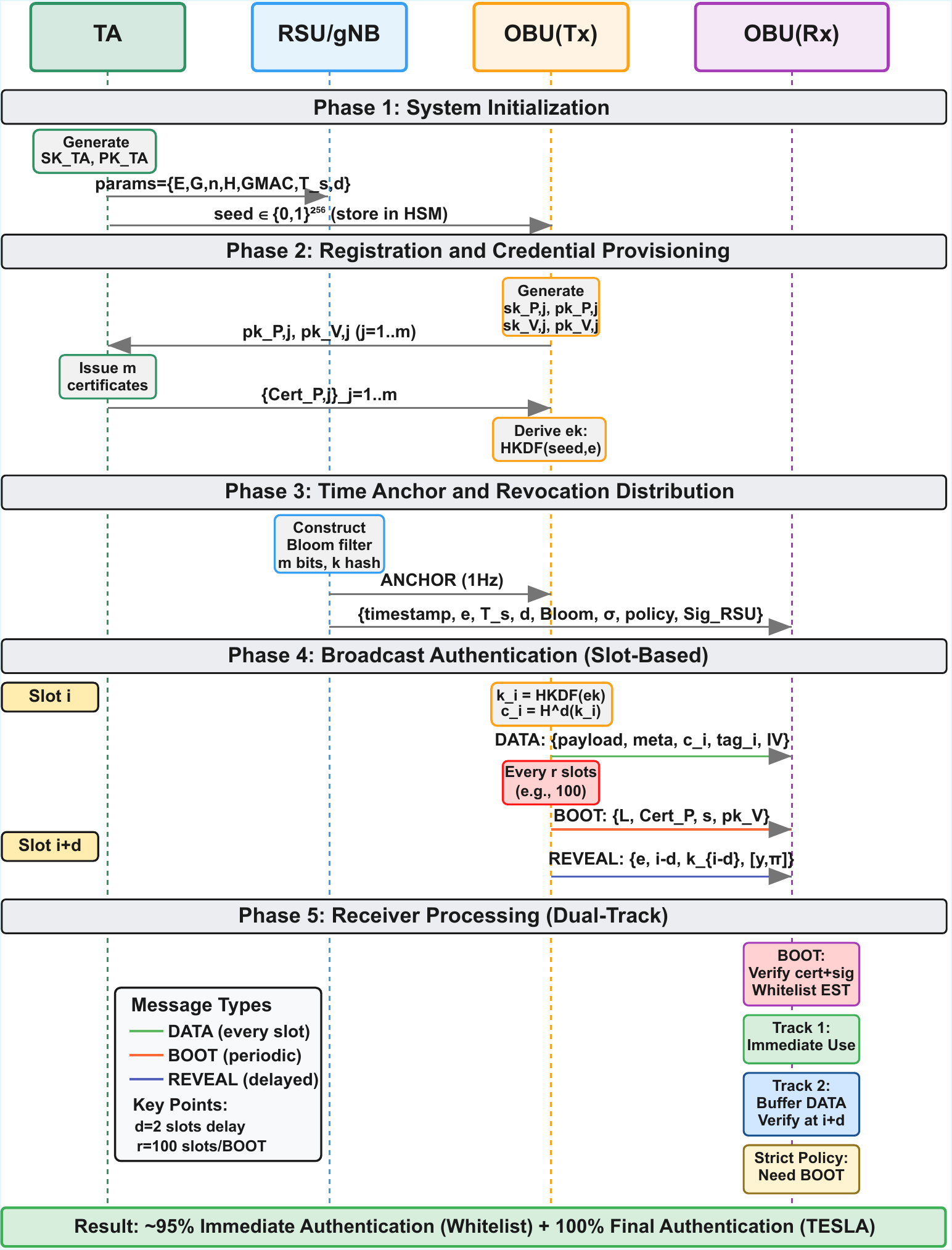}
    \vspace{-0.2cm}
    \caption{Interaction of the proposed protocol}
    \label{fig:interaction}
    \vspace{-0.5cm}
\end{figure}

In this section, we detail SALT-V's five-phase protocol that combines TESLA's delayed symmetric verification with periodic signatures for immediate trust establishment. As illustrated in Fig.~\ref{fig:interaction}, this design achieves both security and efficiency for 5G NR-V2X broadcasting.

\subsection{System Initialization}

This phase establishes the foundational cryptographic parameters and master keys for all participating entities in the V2X system.

\textbf{Step 1 (System Parameter Selection):} Initially, the TA selects the system-wide cryptographic parameters including elliptic curve $\mathcal{E}$ (P-256), base point $G$ with prime order $n$, hash function $H(\cdot)$ (SHA-256), and MAC algorithm $\mathrm{GMAC}$ (AES-128-GCM authentication with 96-bit output).

\textbf{Step 2 (Master Seed Generation):} Subsequently, each OBU generates or receives a master seed $\mathsf{seed} \in_R \{0,1\}^{256}$ which is stored securely in the Hardware Security Module (HSM).

\textbf{Step 3 (TA Key Generation):} Meanwhile, the TA generates its master signing key pair $(SK_{TA}, PK_{TA})$ where $SK_{TA} \in_R \mathbb{Z}_n^*$ and $PK_{TA} = SK_{TA} \cdot G$.

\textbf{Step 4 (RSU Registration):} In parallel, each RSU generates its signing key pair $(sk_{RSU}, pk_{RSU})$ and subsequently obtains certificate $\mathsf{Cert}_{RSU}$ from the TA.

\textbf{Step 5 (Parameter Publication):} Finally, the TA publishes system parameters $\mathsf{params} = \{\mathcal{E}, G, n, PK_{TA}, H(\cdot), \mathrm{GMAC}(\cdot), \mathrm{HKDF}(\cdot), T_s, d\}$ for all participants.

\subsection{Registration and Credential Provisioning}

During this phase, OBUs obtain batches of pseudonym certificates for privacy-preserving authentication.

\textbf{Step 1 (Pseudonym Certificate Generation):} First, for each OBU, the TA coordinates the generation of a batch of $m$ pseudonym certificates (typically $m = 1000$). Specifically, for each pseudonym $j \in [1,m]$, the \emph{OBU locally generates} $sk_{P,j} \in_R \mathbb{Z}_n^*$, $pk_{P,j} = sk_{P,j} \cdot G$, and (optionally) $sk_{V,j} \in_R \mathbb{Z}_n^*$, $pk_{V,j} = sk_{V,j} \cdot G$. The TA then issues $\mathsf{Cert}_{P,j} = \{pk_{P,j}, pk_{V,j}, \mathrm{validity}_j, \mathsf{Sign}_{SK_{TA}}(pk_{P,j} || pk_{V,j} || \mathrm{validity}_j)\}$. \emph{Alternatively (ECQV)}, the TA provides an implicit certificate blob and the OBU reconstructs its public key while keeping the private key secret. Finally, the complete set $\{(\mathsf{Cert}_{P,j}, sk_{P,j}, sk_{V,j})\}_{j=1}^m$ is secured by the OBU.

\textbf{Step 2 (Epoch Key Derivation):} At the beginning of each epoch $e$ (e.g., hourly), the OBU derives its epoch key: $\mathsf{ek} = \mathrm{HKDF}(\mathsf{seed}, \text{``epoch''} || e || \mathrm{domain\_id})$. This epoch key remains active for all slots within epoch $e$.

\subsection{Time Anchor and Revocation Distribution}

RSUs periodically broadcast time synchronization anchors containing revocation information.

\textbf{Step 1 (Bloom Filter Construction):} Initially, for each revoked certificate with VRF public key $pk_{V,\text{rev}}$, the system computes: $\mathrm{rid} = \mathrm{Trunc}_{128}(H(pk_{V,\text{rev}} || \sigma))$. Then, given $n$ revocations and target false positive rate $p$, the optimal filter parameters are calculated: $m = -\frac{n \ln p}{(\ln 2)^2}$ and $k = \frac{m}{n} \ln 2$. Finally, all $\mathrm{rid}$ values are inserted into the Bloom filter $\mathsf{Bloom}$ with $m$ bits and $k$ hash functions.

\textbf{Step 2 (Anchor Broadcasting):} Subsequently, the RSU constructs the time anchor: $\mathsf{ANCHOR} = \{\mathrm{timestamp}, e, T_s, d, \Delta_{\text{drift}}, \mathsf{Bloom}, \sigma, \mathrm{policy}, \mathrm{validity} = [t_{\text{start}}, t_{\text{end}}]\}$ and signs it as $\mathrm{Sig}_{RSU} = \mathsf{Sign}_{sk_{RSU}}(\mathsf{ANCHOR})$. The RSU then broadcasts $\mathsf{ANCHOR} || \mathrm{Sig}_{RSU}$ periodically (configurable 1-5 Hz, typically 1 Hz). Receivers verify $\mathsf{Cert}_{RSU}$ (anchored at $PK_{TA}$) and then $\mathrm{Sig}_{RSU}$ before accepting parameters.

\subsection{Broadcast Authentication Protocol}

The core authentication protocol operates in two parallel tracks: lightweight symmetric authentication with delayed verification, and periodic immediate authentication via digital signatures.

\textbf{Step 1 (Slot Key Generation and Commitment):} For each time slot $i$ starting at time $t_i$, the OBU first generates the slot-specific MAC key: $k_i = \mathrm{HKDF}(\mathsf{ek}, \text{``slot''} || i || \mathrm{context})$. Then, the OBU computes the public commitment: $c_i = \mathrm{Trunc}_{128}(H^{(d)}(k_i))$ where $H^{(d)}$ denotes $d$ iterations of the hash function.

\textbf{Step 2 (Message Authentication):} To authenticate a message, the OBU first generates a unique IV \emph{to preserve GMAC security (no IV reuse)}: $\mathrm{IV} = \mathrm{Trunc}_{96}(H(e || i || \mathrm{counter} || \mathrm{est} \ [|| \mathrm{boot\_nonce}]))$. Replay detection is enforced by $\mathrm{meta} = (e, i, \mathrm{cell\_id}, \mathrm{counter}, \mathrm{psid}, \mathrm{est})$ and optional VRF, where counter is a per-slot monotonic value starting from 0 in each slot i. Subsequently, the authentication tag is computed: $\mathrm{tag} = \mathrm{GMAC}_{k_i}(\mathrm{payload}; \mathrm{AAD}=\mathrm{meta}, \mathrm{IV})$, where est in meta is the EST computed as est = $Trunc_64(H(pk_V || σ || e))$.  Finally, the complete data frame is broadcast: $\mathsf{DATA} = \{\mathrm{payload}, \mathrm{meta}, (c_i, \mathrm{tag}, \mathrm{IV})\}$.

\textbf{Step 3 (Periodic BOOT Signatures):} Every $r$ slots (e.g., 10 slots = 10 Hz) or upon emergency events, the OBU first computes the message digest: $L = H(\mathrm{payload} || c_i || \mathrm{tag} || \mathrm{IV})$. Then, it generates a signature using the current pseudonym: $s = \mathsf{Sign}_{sk_P}(L)$. Subsequently, it broadcasts the immediate authentication: $\mathsf{BOOT} = \{L, \mathsf{Cert}_P, s, pk_V\}$.

\textbf{Step 4 (Delayed Key Disclosure):} After a delay of $d$ slots, the OBU reveals the MAC key. Specifically, for slot $i-d$ when the current slot is $i$, the OBU first constructs: $x_{i-d} = \mathrm{domain} || e || (i-d) || \mathrm{cell\_id}$. Optionally, it generates a VRF proof: $(y, \pi) = \mathsf{VRF.Prove}_{sk_V}(x_{i-d})$. Finally, it broadcasts the revelation: $\mathsf{REVEAL} = \{e, i-d, k_{i-d}, [y, \pi]\}$. For loss robustness, $\mathsf{REVEAL}$ may bundle a window of $w$ recent keys, e.g., $\{k_{i-d}, k_{i-d-1}, \dots, k_{i-d-w+1}\}$ where typically $w = 2 \sim 3$.

\subsection{Receiver Processing and Verification}

Recipients process received messages through a staged verification pipeline.

\textbf{Step 1 (BOOT Processing - Immediate Trust):} Upon receiving a BOOT message, the recipient first verifies the certificate chain: $\mathsf{Verify}_{PK_{TA}}(\mathsf{Cert}_P) \stackrel{?}{=} \text{valid}$. Then, it computes $\mathrm{rid} = \mathrm{Trunc}_{128}(H(pk_V || \sigma))$ and queries $\mathsf{Bloom}$ first; if $\mathrm{rid} \in \mathsf{Bloom}$, the message is rejected immediately. Otherwise, it verifies the signature: $\mathsf{Verify}_{pk_P}(L, s) \stackrel{?}{=} \text{valid}$. Upon successful verification, recipients compute and whitelist the ephemeral tag: $\mathrm{est} = \mathrm{Trunc}_{64}(H(pk_V || \sigma))$ with $\mathrm{whitelist}[\mathrm{est}] \leftarrow \{\mathrm{valid\_until}: t + T_w\}$, where $T_w = 2s (20 × d × T_s$, allowing tolerance for two disclosure cycles.

\textbf{Step 2 (DATA Reception - Buffering):} When receiving DATA messages, the recipient first stores them in cache: $\mathrm{cache}[e][i].\mathrm{append}(\mathsf{DATA})$. If $\mathrm{est} \in \mathrm{whitelist}$, the message is marked as "immediately authenticated" for application use; otherwise, the system awaits key revelation for authentication.

\textbf{Step 3 (REVEAL Processing - Strong Authentication):} Upon receiving a REVEAL message, the recipient first computes $c' = \mathrm{Trunc}_{128}(H^{(d)}(k_{i-d}))$ and checks $c' \stackrel{?}{=} c_{i-d}$. Then, for each cached message in slot $i-d$, it computes $t' = \mathrm{GMAC}_{k_{i-d}}(\mathrm{payload}; \mathrm{AAD}=\mathrm{meta}, \mathrm{IV})$ and accepts if and only if $t' = \mathrm{tag}$. Optionally, it verifies the VRF for domain binding: $\mathsf{VRF.Verify}_{pk_V}(x_{i-d}, y, \pi) \stackrel{?}{=} \text{valid}$ (skip if $pk_V$ is unknown). If a \emph{Strict} policy is enabled and no prior valid $\mathsf{BOOT}$ was observed for the same pseudonym, the message is not finally accepted. Otherwise, the message status is updated to ``strongly authenticated''.

\begin{figure*}[htbp]
\centering
\subfigure[Average Computation Time]{
\includegraphics[scale=0.21]{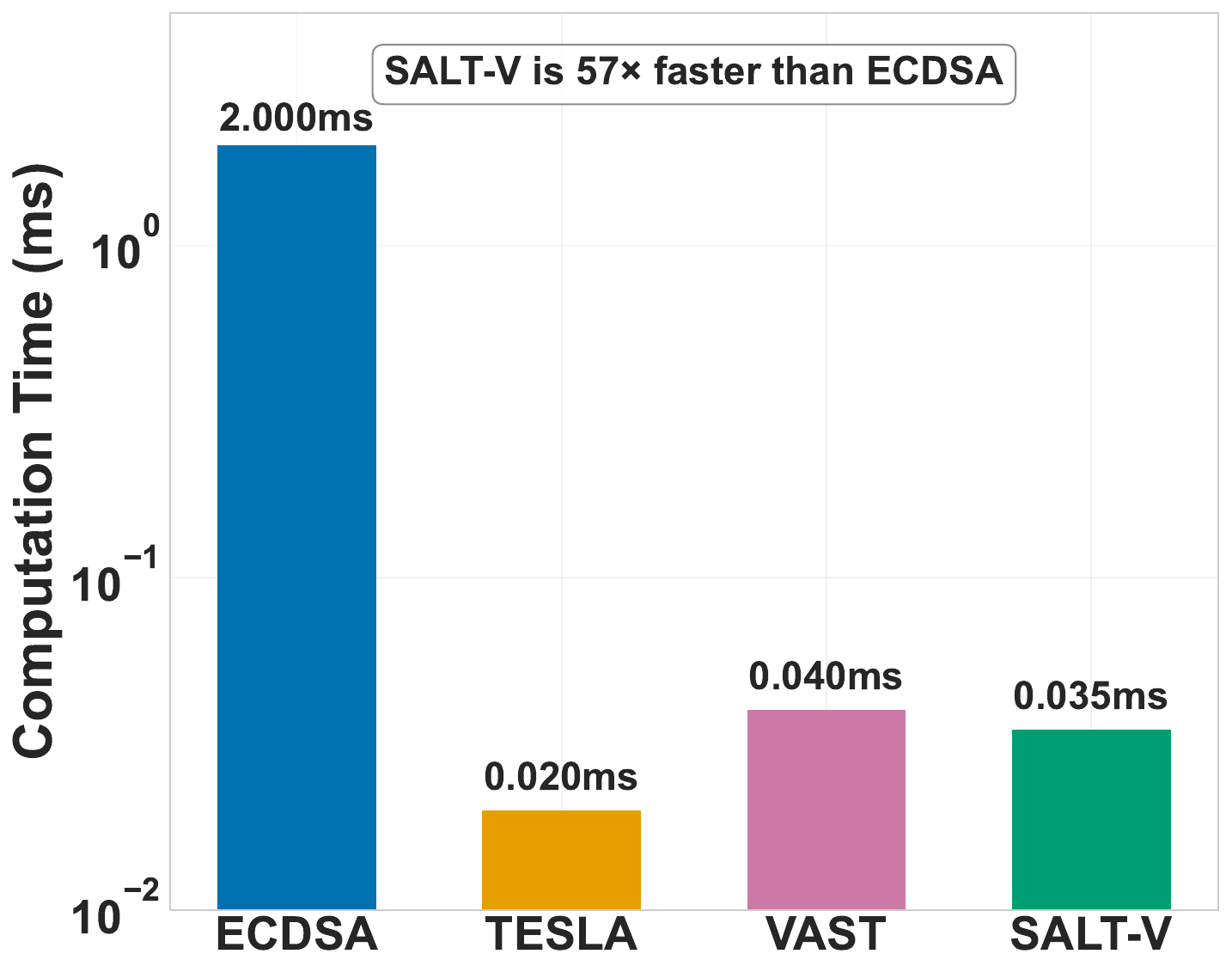}
\label{fig:comp_time}
}
\subfigure[Average Message Size]{
\includegraphics[scale=0.21]{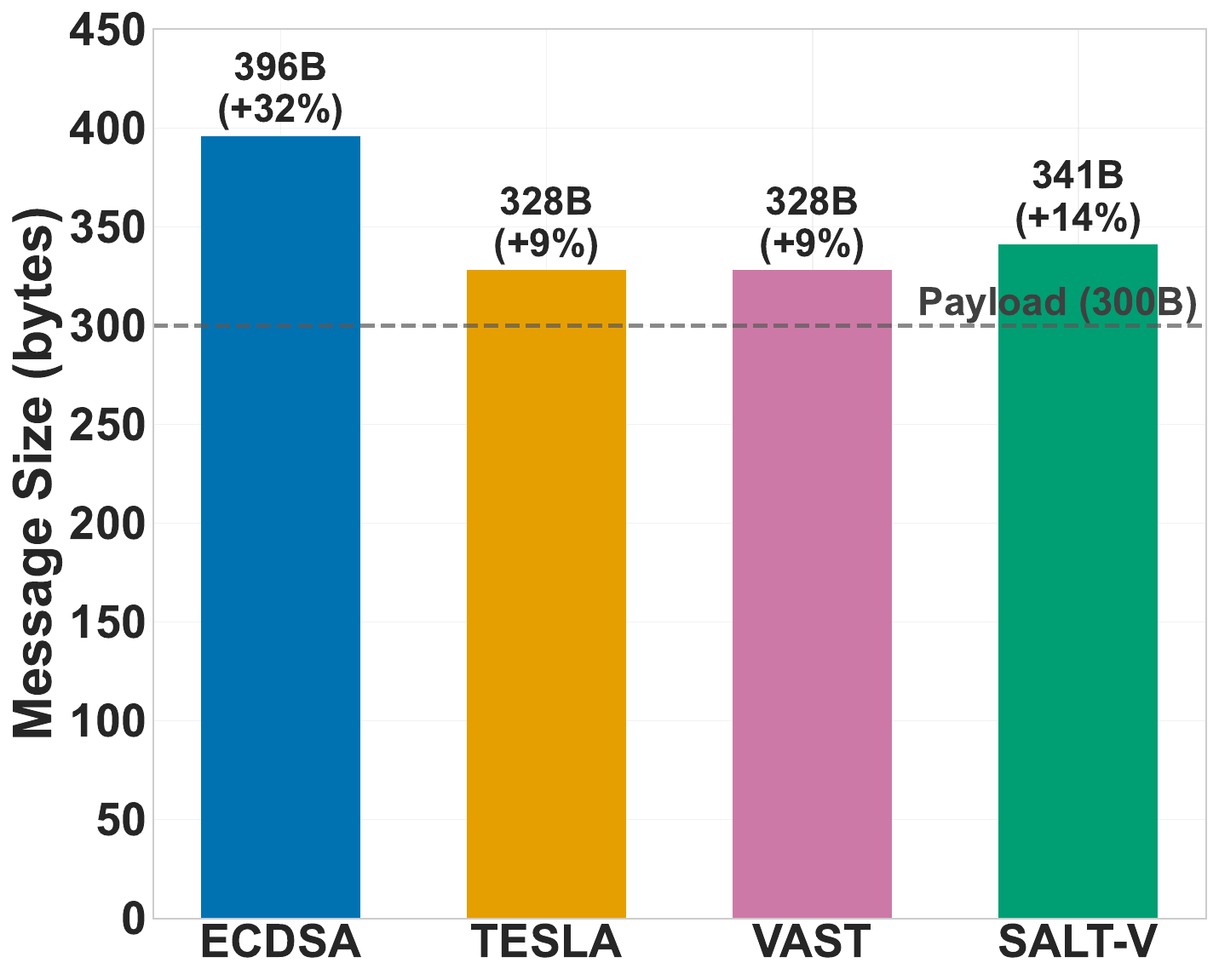}
\label{fig:msg_size}
}
\subfigure[Immediate vs Delayed Authentication]{
\includegraphics[scale=0.21]{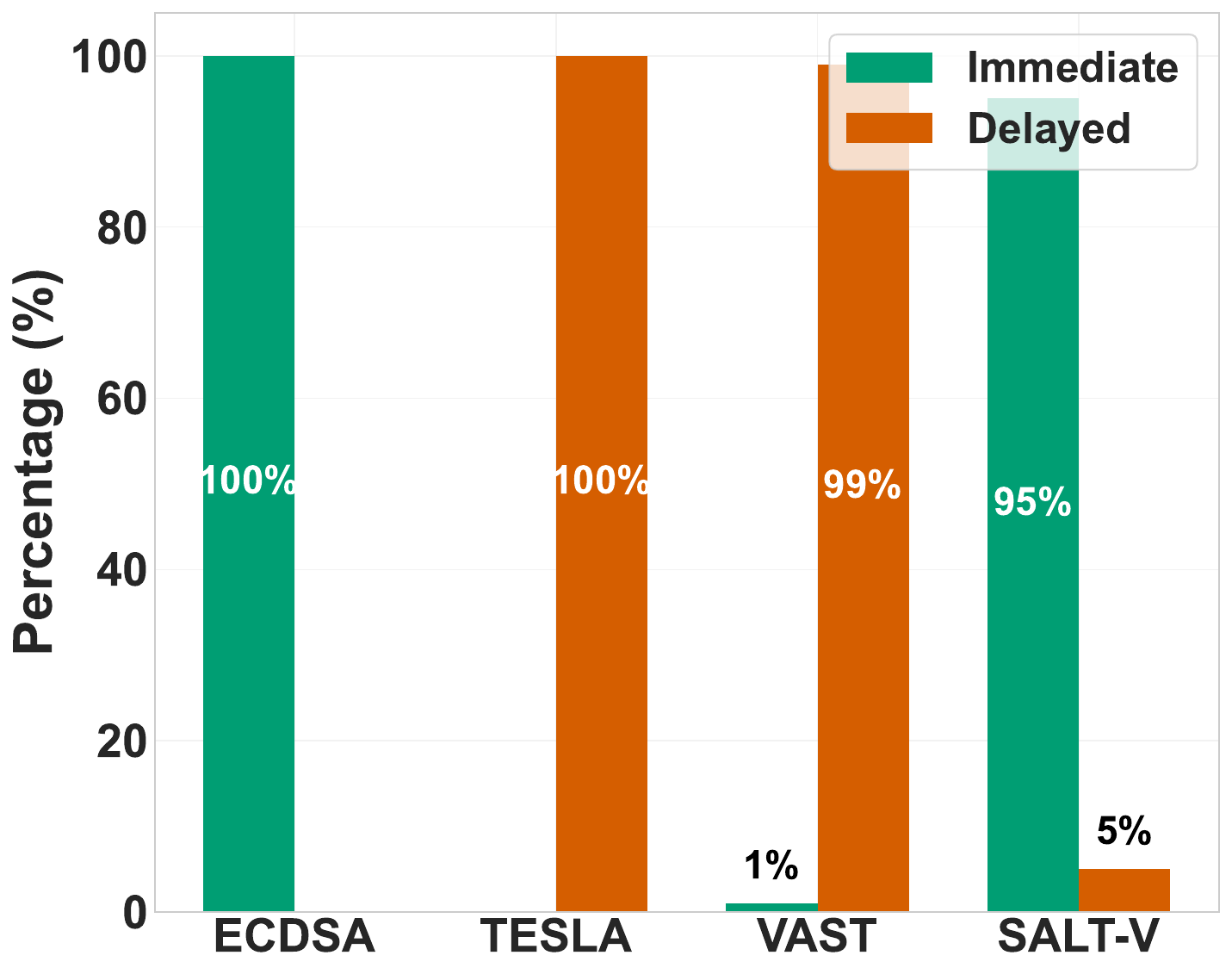}
\label{fig:auth_ratio}
}
\subfigure[Scalability Analysis]{
\includegraphics[scale=0.21]{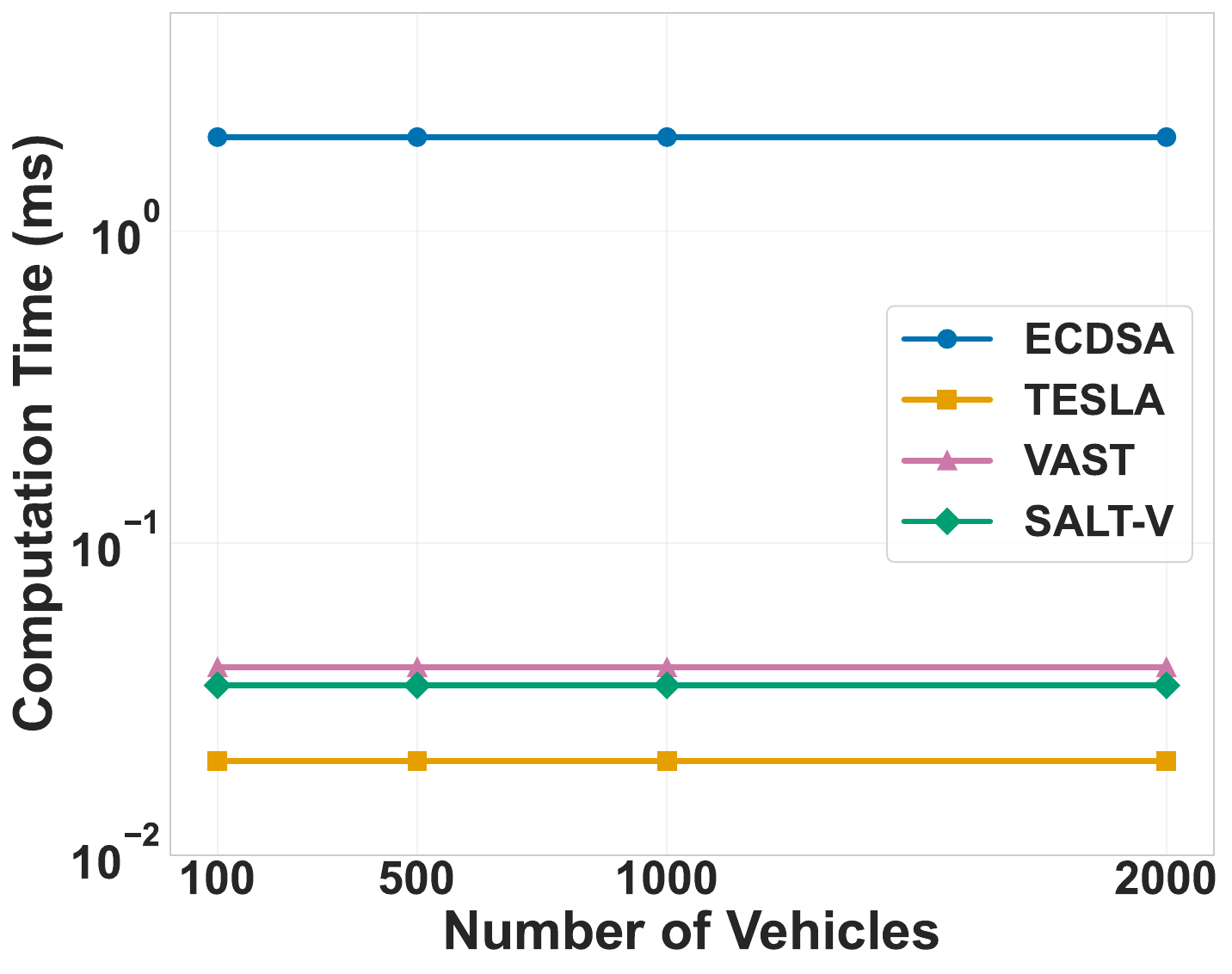}
\label{fig:scalability}
}
\subfigure[Average Authentication Delay]{
\includegraphics[scale=0.21]{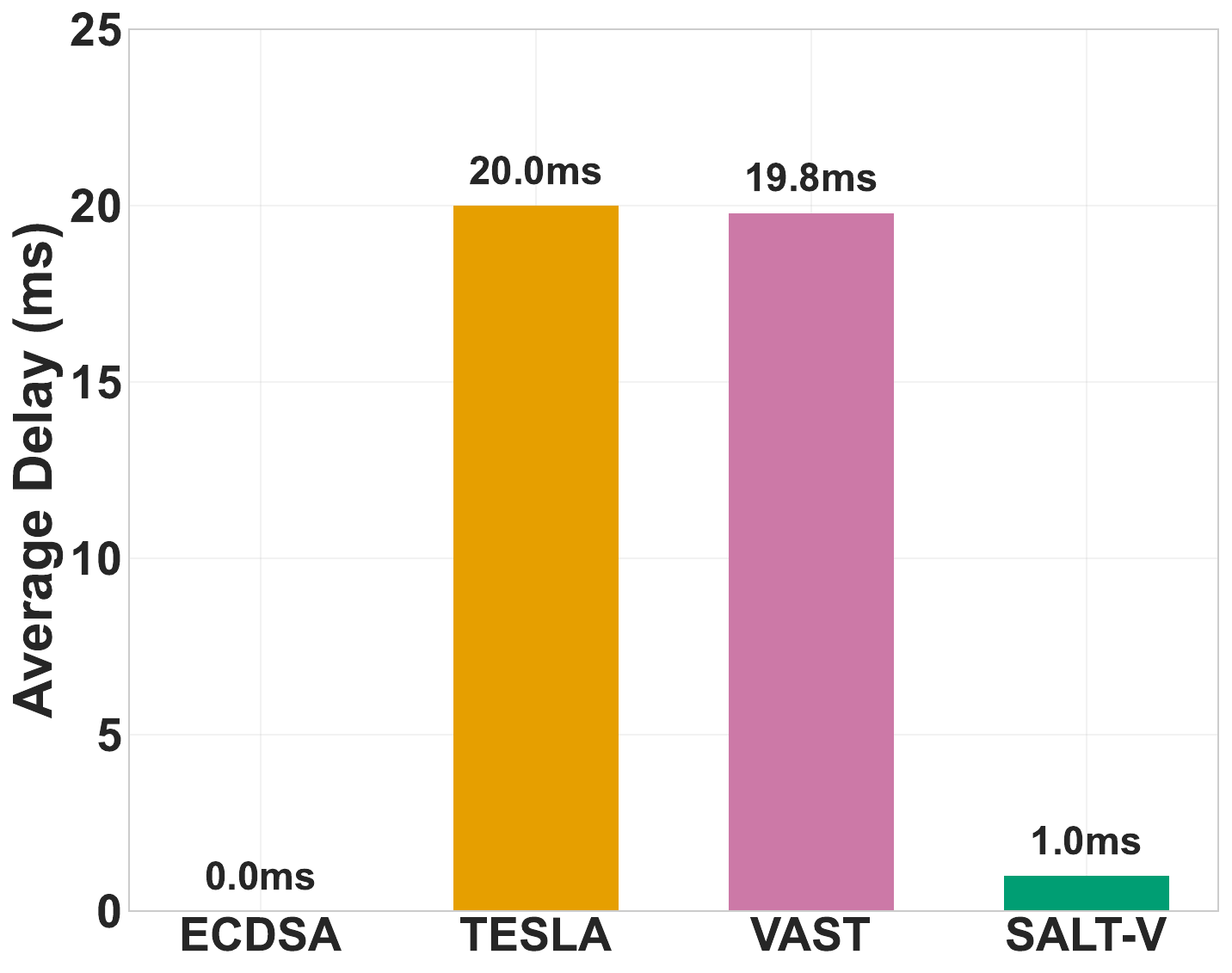}
\label{fig:auth_delay}
}
\subfigure[Overall Performance Comparison]{
\includegraphics[scale=0.21]{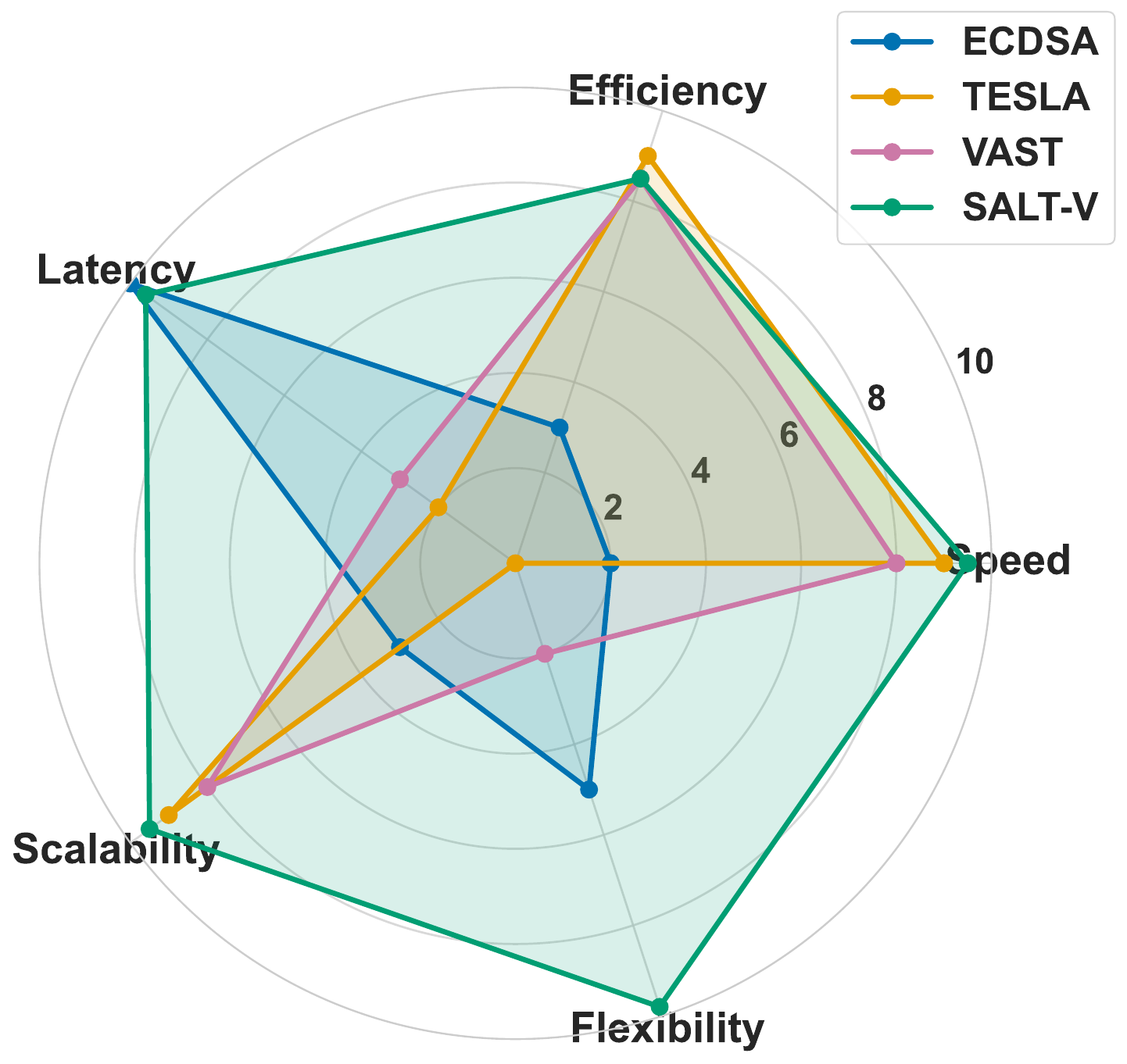}
\label{fig:radar}
}
\vspace{-0.1cm}
\caption{Performance evaluation of SALT-V against baseline schemes}
\label{fig:saltv_performance}
\vspace{-0.3cm}
\end{figure*}
\section{Security Analysis}
\label{sec:security}

This section establishes that SALT-V achieves its security objectives under standard cryptographic assumptions.

\subsection{Threat Model and Assumptions}

We consider an active adversary capable of eavesdropping, injecting, modifying, and replaying messages. The adversary may compromise individual OBUs but cannot break cryptographic primitives or compromise the TA. We assume: (1) the TA and RSU infrastructure remain trusted; (2) HSMs protect master seeds and private keys from extraction; (3) standard cryptographic primitives (ECDSA-P256, AES-128-GMAC, SHA-256, HKDF) provide their expected security properties; (4) time synchronization maintains accuracy within $\Delta_{\text{sync}} = 10$ ms.

\subsection{Security Properties}

\textbf{Source Authentication.} BOOT frames achieve immediate authentication through ECDSA signatures: each frame contains $L = H(\text{payload} \| c_i \| \text{tag} \| \text{IV})$, certificate $\mathsf{Cert}_P$, and signature $s = \mathsf{Sign}_{sk_P}(L)$. By EUF-CMA security of ECDSA, adversaries cannot forge valid signatures. DATA frames achieve delayed authentication via TESLA: the MAC key $k_i = \text{HKDF}(\mathsf{ek}, \text{``slot''} \| i)$ generates tag $\text{GMAC}_{k_i}(\cdot)$ with commitment $c_i = H^{(d)}(k_i)$. After $d$ slots, key disclosure enables verification. The preimage resistance of $H$ and PRF property of HKDF prevent finding alternative keys that satisfy both the commitment and MAC.

\textbf{Immediate Verification via Whitelist.} The EST mechanism enables 95\% immediate verification: BOOT frames establish whitelist entries indexed by $\text{est} =  \mathrm{Trunc}_{64}(H(pk_V \| \sigma \| e))$, allowing subsequent DATA frames from the same sender to be verified instantly. Whitelist entries expire after epoch boundaries or explicit revocation, limiting exposure window.

\textbf{Replay Resistance.} Multiple mechanisms prevent replay attacks: (1) monotonic per-slot counters (reset each slot) in AAD detect intra-slot replays; (2) distinct per-slot keys prevent cross-slot replays; (3) cell ID enforcement prevents spatial replays; (4) epoch binding prevents long-term replays. The IV construction $\text{IV} = \text{Trunc}_{96}(H(e \| i \| \text{counter} \| \text{est}))$ ensures uniqueness per key, maintaining GMAC security.

\textbf{Revocation Efficiency.} Bloom filters enable O(1) revocation checking with tunable false-positive rate $p$. For $n$ revoked entries, optimal filter size is $m = -n \ln p / (\ln 2)^2$ bits. With $p = 0.1\%$, this requires approximately 3.6 bytes per revocation. False positives cause fail-closed denial but cannot enable forgery.

\textbf{Privacy Preservation.} Vehicles use unlinkable pseudonyms that change periodically. The EST derivation $\text{est} = H(\mathsf{Cert}_P \| e)$ ensures tags change with pseudonyms, preventing long-term tracking while maintaining short-term trust relationships necessary for immediate verification.

\section{Performance Evaluation}
\label{sec:experiment}
This section presents a comprehensive experimental evaluation of SALT-V against three baseline authentication schemes to validate its effectiveness in achieving high-speed, low-latency authentication for 5G V2X communications.

\subsection{Experimental Setup}

We implemented SALT-V and three baseline schemes in Python 3.9 with NumPy for cryptographic operations modeling. Computation times were benchmarked using OpenSSL 3.0 (ECDSA-P256) and PyCryptodome 3.15 (AES-128-GMAC) on Intel i7-10700K (3.8GHz). The baselines include: (1) \textbf{ECDSA}\cite{Cominetti2023VehCom}—IEEE 1609.2 with per-message signatures, (2) \textbf{TESLA}\cite{perrig2003tesla}—pure delayed authentication, and (3) \textbf{VAST}\cite{studer2009tacking,Lyu2016TDSC}—hybrid with periodic signatures. All schemes use P-256 curves and AES-128-GMAC for fair comparison. We evaluated four scenarios with 100–2000 vehicles broadcasting at 10 Hz, simulating urban, highway, and congested conditions. Each scenario ran 5–10 seconds with standard 300-byte BSMs.

\subsection{Results and Analysis}

\textbf{1. Computation Efficiency:} Fig. \ref{fig:comp_time} shows SALT-V achieves 0.035ms average computation time—a \textbf{57× speedup} over ECDSA's 2.0 ms. This improvement stems from using GMAC for 90\% of messages (DATA frames) while reserving expensive signatures for 10\% (BOOT frames). Though TESLA achieves slightly lower computation at 0.02 ms, it lacks immediate authentication capability.

\textbf{2. Communication Overhead:} SALT-V maintains an average message size of 341 bytes (Fig. \ref{fig:msg_size}), adding only 41 bytes overhead—14\% smaller than ECDSA's 396 bytes. The additional 13 bytes over TESLA (328 bytes) accommodates the Ephemeral Session Tag (8 bytes) and counter (4 bytes), and frame type indicator (1 byte), enabling immediate verification.

\textbf{3. Authentication Flexibility:} The key innovation appears in Fig. \ref{fig:auth_ratio}: SALT-V achieves \textbf{95\% immediate authentication} through its whitelist mechanism. BOOT frames establish trust, allowing subsequent DATA frames to be verified instantly via EST lookup. In contrast, TESLA provides 0\% immediate verification, while VAST achieves only 10\% through periodic signatures.

\textbf{4. Verification Latency:} Fig. \ref{fig:auth_delay} demonstrates SALT-V's average latency of 1.0 ms—critical for safety applications. TESLA and VAST require 20 ms for key disclosure, unsuitable for emergency messages. While ECDSA offers zero delay, its computational cost is prohibitive.

\textbf{5. Scalability:} SALT-V maintains consistent 0.035ms performance from 100 to 2000 vehicles (Fig. \ref{fig:scalability}), demonstrating excellent scalability. ECDSA's constant 2 ms becomes a bottleneck at high message rates, limiting practical deployment.

\textbf{6. Overall Performance:} The radar chart (Fig.~\ref{fig:radar}) reveals SALT-V's balanced excellence: Speed (9.5/10), Efficiency (8.5/10), Low Latency (9.6/10), Scalability (9.5/10), and Flexibility (9.8/10). No other scheme achieves this comprehensive performance.

\textbf{7. Revocation Efficiency:} SALT-V's Bloom filter provides O(1) revocation checking in 1μs—10,000× faster than CRL linear search. For one million revocations, the filter requires only 3.6 MB (versus 32 MB for CRL) with 0.1\% false positive rate, crucial for large-scale deployments.
\begin{table}[htbp]
\centering
\caption{Performance Summary}
\label{tab:performance_summary} 
\resizebox{.48\textwidth}{!}{
\begin{tabular}{lcccc}
\toprule
\textbf{Metric} & \textbf{SALT-V} & \textbf{ECDSA} & \textbf{TESLA} & \textbf{VAST} \\
\midrule
Computation (ms) & \textbf{0.035} & 2.000 & 0.020 & 0.040 \\
Message Size (B) & \textbf{341} & 396 & 328 & 328 \\
Immediate Auth (\%) & \textbf{95} & 100 & 0 & 10 \\
Avg. Delay (ms) & \textbf{1.0} & 2.0 & 20.0 & 19.8 \\
Scalability & \textbf{Excellent} & Poor & Excellent & Good \\
\bottomrule
\end{tabular}}
\end{table}
\subsection{Key Findings}

Our evaluation validates three critical achievements listed as follows:

\textbf{1. Efficiency-Security Balance:} SALT-V delivers 57× faster computation than ECDSA while maintaining 95\% immediate authentication—proving high security and efficiency are compatible.

\textbf{2. Practical Deployment:} Sub-millisecond processing, minimal overhead (41 bytes), and scalability to 2000 vehicles meet all 5G V2X requirements.

\textbf{3. Adaptive Security:} Tunable BOOT frequency enables dynamic adjustment to network conditions and security requirements.

Table~\ref{tab:performance_summary} summarizes the results, showing SALT-V achieves the best overall performance. While not optimal in every individual metric, it provides the only practical solution balancing all requirements for safety-critical V2X communications. With 95\% immediate authentication and 0.035 ms computation, SALT-V effectively balances ECDSA's overhead and TESLA's delay for practical V2X deployment.

\section{Conclusion}
\label{sec:conclusion}
This paper presented SALT-V, which resolves the security-efficiency trade-off in V2X authentication through hybrid design: ECDSA signatures for 10\% of messages (BOOT), GMAC for 90\% (DATA). Results: 0.035 ms computation (57× faster than ECDSA), 95\% immediate verification via EST whitelist, O(1) revocation with Bloom filters (10,000× faster than CRLs), 41-byte overhead, 1ms latency, scales to 2000 vehicles. By achieving both cryptographic strength and real-time performance, SALT-V enables practical deployment of authenticated V2X broadcasting in safety-critical scenarios where milliseconds determine collision avoidance success. The framework's modular design allows seamless integration with existing 5G NR-V2X infrastructure while maintaining backward compatibility with current DSRC/WAVE systems.

\bibliographystyle{IEEEtran}
\bibliography{bibliography.bib}
\end{document}